\documentclass[sigconf]{acmart}

\AtBeginDocument{%
  }

\usepackage{enumitem}
\usepackage{graphicx}
\usepackage{caption}
\usepackage{subcaption}
\usepackage{lipsum} 
\usepackage{amsmath}
\usepackage{amsfonts}
\usepackage{amsthm}

\usepackage{multirow}
\usepackage{multicol}

\usepackage{todonotes}
\usepackage{textcomp}
\usepackage{subcaption}
\usepackage{threeparttable}

\setcopyright{acmlicensed}
\copyrightyear{2025}
\acmYear{2025}
\acmDOI{XXXXXXX.XXXXXXX}

\acmISBN{978-1-4503-XXXX-X/18/06}

\setcopyright{none}
\begin{document}

\title{Generative Sequential Recommendation via Hierarchical Behavior Modeling}


\author{Zhefan Wang}
\authornote{Both authors contributed equally to this research.}
\affiliation{
  \institution{DCST, Tsinghua University}
  \city{Beijing}
  \country{China}
}
\email{wzf23@mails.tsinghua.edu.cn}

\author{Guokai Yan}
\authornotemark[1]
\affiliation{
  \institution{Kuaishou Technology}
  \city{Beijing}
  \country{China}
}
\email{yanguokai@kuaishou.com}

\author{Jinbei Yu}
\affiliation{
  \institution{Kuaishou Technology}
  \city{Beijing}
  \country{China}
}
\email{yujinbei@kuaishou.com}

\author{Siyu Gu}
\authornote{Corresponding author.}
\affiliation{
  \institution{Kuaishou Technology}
  \city{Beijing}
  \country{China}
}
\email{gusiyu@kuaishou.com}

\author{Jingyan Chen}
\affiliation{
  \institution{Kuaishou Technology}
  \city{Beijing}
  \country{China}
}
\email{chenjingyan03@kuaishou.com}

\author{Peng Jiang}
\affiliation{
  \institution{Kuaishou Technology}
  \city{Beijing}
  \country{China}
}
\email{jiangpeng@kuaishou.com}

\author{Zhiqiang Guo}
\affiliation{%
  \institution{DCST, Tsinghua University}
  \city{Beijing}
  \country{China}}
\email{georgeguo.gzq.cn@gmail.com}

\author{Min Zhang}
\authornotemark[2]
\affiliation{%
  \institution{DCST, Tsinghua University}
  \city{Beijing}
  \country{China}}
\email{z-m@tsinghua.edu.cn}


%

\keywords{Generative Recommendation, Sequential Recommendation, Multi-Behavior Modeling, Sequential Augmentation}



\begin{abstract}


Recommender systems in multi-behavior domains, such as advertising and e-commerce, aim to guide users toward high-value but inherently sparse conversions.
Leveraging auxiliary behaviors (e.g., clicks, likes, shares) is therefore essential.
Recent progress on generative recommendations has brought new possibilities for multi-behavior sequential recommendation.
However, existing generative approaches face two significant challenges: 
1) \textbf{Inadequate Sequence Modeling}: capture the complex, cross-level dependencies within user behavior sequences,
and 2) \textbf{Lack of Suitable Datasets}: publicly available multi-behavior recommendation datasets are almost exclusively derived from e-commerce platforms, limiting the validation of feasibility in other domains, while also lacking sufficient side information for semantic ID generation.

To address these issues, we propose a novel generative framework, \textbf{GAMER} (\textbf{G}enerative \textbf{A}ugmentation and \textbf{M}ulti-l\textbf{E}vel behavior modeling for \textbf{R}ecommendation), built upon a decoder-only backbone.
GAMER introduces a cross-level interaction layer to capture hierarchical dependencies among behaviors and a sequential augmentation strategy that enhances robustness in training.
To further advance this direction, we collect and release \textbf{ShortVideoAD}, a large-scale multi-behavior dataset from a mainstream short-video platform, which differs fundamentally from existing e-commerce datasets and provides pretrained semantic IDs for research on generative methods.
Extensive experiments show that GAMER consistently outperforms both discriminative and generative baselines across multiple metrics.\footnote{Code and data are available at \url{https://github.com/wzf2000/GAMER}.}

\end{abstract}

\maketitle


\section{Introduction}

Recommender systems serve as critical infrastructure in numerous online platforms, ranging from e-commerce sites to streaming services.
In many scenarios, particularly in advertising and e-commerce, user interactions naturally form a multi-behavioral hierarchy.
While platforms primarily optimize for deeper, conversion-oriented behaviors such as purchases or ad conversions, these key signals are inherently sparse (e.g., conversions appear less than 4 times on average in each user's recent history).
However, this sparsity results in individual behavior sequences that are too short and fragmented to support meaningful sequential modeling.
In contrast, auxiliary behaviors (e.g., clicks and likes) occur more frequently, forming a rich behavioral pathway that potentially reveals nuanced user intent.
Effectively leveraging this entire spectrum of multi-behavior historical sequences presents a significant opportunity to alleviate data sparsity and enhance recommendation accuracy.

Recently, generative recommendation paradigms have opened new directions for sequential recommendation.
By reframing recommendations as a sequence generation task, generative methods have demonstrated remarkable capabilities in modeling long-range dependencies and capturing complex user preferences.
This paradigm is particularly appealing for multi-behavior recommendation, where user interactions form hierarchical sequences.

However, when applied to the multi-behavior setting, existing generative methods face several critical challenges:
1) \textbf{Inadequate Sequence Modeling}: Most approaches either treat different behaviors as independent sequences or concatenate them, without explicitly capturing the hierarchical dependencies among behavior types.
As a result, they fail to fully exploit the rich structural information embedded in multi-level user interactions.
2) \textbf{Lack of Suitable Datasets}: Publicly available multi-behavior datasets are predominantly collected from e-commerce platforms~\cite{kim2025multi}, which contain large numbers of repetitive interactions and product redirections that differ substantially from the interaction dynamics in short-video advertising.
Moreover, most datasets lack side information that could provide semantically meaningful identifiers for items, limiting their utility for generative recommendation research that benefits from rich semantic representations.

To address these challenges, a novel generative framework has been proposed in this work with the following contributions:

\begin{itemize}
  \item We propose \textbf{GAMER}, a novel multi-behavior generative recommendation framework built on a decoder-only architecture.
  GAMER introduces a dedicated cross-level behavior interaction layer and a sequential augmentation strategy, enabling explicit hierarchical modeling and improved robustness to rare behaviors.
  \item We collect and release \textbf{ShortVideoAD}, a large-scale multi-behavior dataset from a mainstream short-video platform. Unlike existing e-commerce datasets, ShortVideoAD captures diverse, ephemeral ad-related behaviors and additionally provides pretrained semantic IDs, offering a valuable testbed for future generative recommendation research.
  \item Extensive experiments on both ShortVideoAD and two public e-commerce multi-behavior recommendation datasets demonstrate the effectiveness of our proposed GAMER, with particularly significant improvements observed in the short-video advertising scenario.
  Further experiments demonstrate the generality of our proposed sequential augmentation method. 
\end{itemize}

\section{Related Work}

\subsection{Generative Recommendation}

Generative recommendation has recently emerged as a promising paradigm that unifies the retrieval and ranking pipeline into a single generative process.
A representative work is TIGER~\cite{rajput2023recommender}, which introduces semantic item identifiers and trains an auto-regressive model to directly generate the identifier of the next item based on user history.
This approach effectively bypasses traditional embedding-based retrieval and demonstrates advantages in cold-start and generalization scenarios.
Building upon this paradigm, several extensions have been proposed.
EAGER~\cite{wang2024eager} designs a two-stream architecture that decodes behavior and semantic signals separately, encouraging collaboration between the two modalities.
PinRec~\cite{badrinath2025pinrec} further explores industrial-scale deployment by conditioning the generation process on different user outcomes and allowing multi-token outputs to improve diversity.
Meanwhile, OneRec~\cite{deng2025onerec,zhou2025onerec,zhou2025onerecv2} unifies retrieval and ranking by generating session-level candidate sequences with an iterative preference alignment mechanism.  
Similarly, UniGRF~\cite{zhang2025killing} introduces a ranking-driven enhancer to bridge the retrieval and ranking objectives within a single generative framework.

Beyond item-level generation, recent efforts have begun to consider multi-behavior sequential recommendation.
MBGen~\cite{liu2024multi} formulates the task as interleaved token generation, where the model auto-regressively generates both user behavior types and item semantic tokens in a unified sequence.
This design enables joint modeling of heterogeneous interactions and provides stronger supervision for learning user preferences.

Compared with these approaches, our work further advances generative recommendation in the multi-behavior scenario.
In particular, we design multi-behavior sequential augmentations, which distinguish our model from prior generative recommenders.

\subsection{Multi-Behavior Recommendation}

User interactions on many real-world platforms often involve multiple types of feedback, including clicks, purchases, ratings, and likes.
To fully capture user preferences, early studies extended traditional collaborative filtering frameworks to jointly model heterogeneous feedback, such as collective matrix factorization or tensor factorization~\cite{rendle2010factorization,karatzoglou2010multiverse}.
With the advent of deep learning, neural models were introduced to capture non-linear correlations among behaviors.  
Representative works include NMTR~\cite{gao2019neural}, which leverages neural collaborative filtering to model dependencies across feedback types, and MBGCN~\cite{jin2020multi}, which employs graph neural networks to learn behavior-specific embeddings while exploiting cross-behavior correlations.

Another important line of research focuses on sequential multi-behavior recommendation.
These methods aim to capture the dynamic dependencies across different types of feedback over time.
More recent works explore transformer-based architectures that differentiate item and behavior tokens in auto-regressive modeling~\cite{yang2022multi}.
In addition, there has been an emerging interest in adopting generative paradigms to unify the modeling of heterogeneous behaviors within sequential frameworks~\cite{liu2024multi,tan2025implicit}.
These developments demonstrate the growing importance of integrating heterogeneous feedback into both static and dynamic recommendation models.
Compared with these studies, our work further advances this line of research by introducing multi-level behavior modeling and behavior-aware sequential augmentations.

\subsection{Data Augmentation for Sequential Recommendation}

Data augmentation techniques in sequential recommendation are primarily designed to address data sparsity and enhance model generalization by creating diverse training samples.
Some researchers directly manipulate the raw item sequences to generate new training examples.
Common techniques include sliding window~\cite{tang2018personalized,zhou2024contrastive}, cropping~\cite{xie2022contrastive}, reordering~\cite{xie2022contrastive}, masking~\cite{sun2019bert4rec,tan2016improved}, substitution~\cite{liu2021contrastive}, and insertion~\cite{liu2021contrastive}.
These methods are computationally efficient and widely adopted in contrastive learning frameworks to generate positive views for self-supervised training.
Besides, some studies leverage advanced neural architectures to generate high-quality synthetic sequences.
L2Aug~\cite{wang2022learning} learns an augmentation policy to generate synthetic sequences from core user data.
SSDRec~\cite{zhang2024ssdrec} performs self-augmentation and applies hierarchical denoising to remove noise and refine the sequence.

However, in multi-behavior recommendation scenarios, user sequences encompass not only the items interacted with but also the types of user behavior.
In this work, we enhance the performance of multi-behavior recommender systems through behavior-aware sequential augmentation.

\begin{figure*}[ht]
    \includegraphics[width=\textwidth]{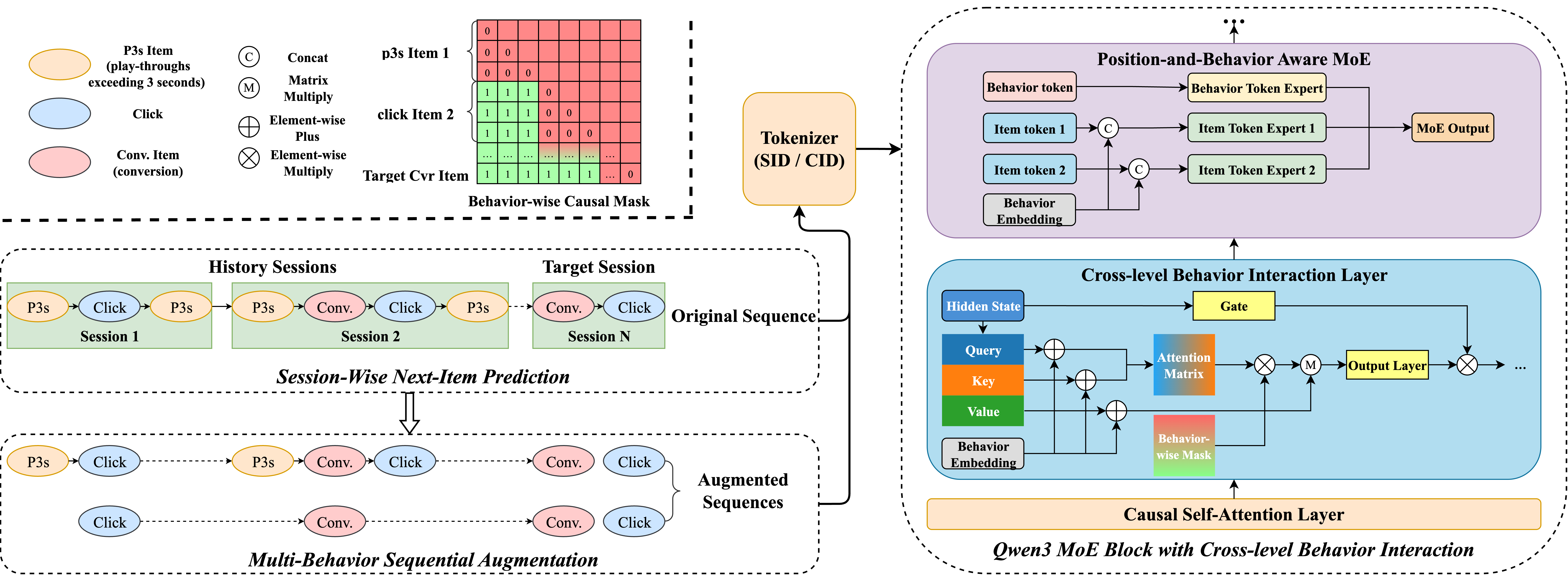}
    \caption{The overview of our proposed GAMER. The left illustration shows our multi-behavior sequential augmentation, which samples the original sequence with different dropout rates $r^t, t=1, \ldots, x$ to generate $x$-fold additional training samples. The right illustrates our Qwen3 MoE block, which consists of three modules: Causal Self-Attention Layer, Cross-level Behavior Interaction Layer, and Position-and-Behavior Aware MoE.}
    \Description{}
    \label{fig:framework}
\end{figure*}

\section{Methodology: GAMER}

In this section, we present the proposed multi-behavior generative recommendation algorithm, \textbf{G}enerative \textbf{A}ugmentation and \textbf{M}ulti-l\textbf{E}vel behavior modeling for \textbf{R}ecommendation (\textbf{GAMER}).
To ensure a rigorous evaluation, we adopt a session-wise evaluation protocol (Section~\ref{sec:session-wise}), a decision motivated by an analysis of the limitations inherent in conventional evaluation protocols for multi-behavior recommendation.
Our methodology primarily consists of two key components: multi-behavior sequential augmentation and Qwen3 MoE block with cross-level behavior interaction, which are elaborated upon in Section~\ref{sec:augmentation} and Section~\ref{sec:cross}, respectively. The overall framework of the proposed model is illustrated in Figure~\ref{fig:framework}.

\subsection{Task Formulation}

In this paper, we formalize the multi-behavior sequential recommendation task under a session-aware setting.
We denote the set of all users as $\mathcal{U} = \{u_1, \cdots, u_{|\mathcal{U}|}\}$ and the set of all items as $\mathcal{V} = \{v_1, v_2, \cdots, v_{|\mathcal{V}|}\}$.
The set of behavior types (e.g., play, click, conversion) is denoted as $\mathcal{B} = \{b_1, b_2, \cdots, b_{|\mathcal{B}|}\}$.
The training set, validation set, and test set are denoted as $\mathcal{D}^{\rm{train}}, \mathcal{D}^{\rm{val}}, \mathcal{D}^{\rm{test}}$, respectively.

Each user $u \in \mathcal{U}$ is associated with a chronologically ordered interaction sequence $\mathcal{S}_u = [(b_1, v_1), (b_2, v_2), \dots, (b_n, v_n)]$,
where each tuple $(b_i, v_i)$ indicates that user $u$ interacted with item $v_i \in \mathcal{V}$ through behavior type $b_i \in \mathcal{B}$.\footnote{$v_i$ contains possible repeated interaction items.}
The sequence $\mathcal{S}_u$ is further partitioned into sessions based on a predefined criterion (e.g., time gap).
Let the session set of user $u$ be $\{\pmb{S}_1, \pmb{S}_2, \dots, \pmb{S}_m\}$, where each session $\pmb{S}_s$ is a contiguous subsequence of $\mathcal{S}_u$.

\subsection{Session-Wise Next Item Prediction}
\label{sec:session-wise}

In the context of multi-behavior sequential recommendation, user behaviors can be categorized into multiple levels based on their depth of engagement.
For instance, in e-commerce recommendation scenarios, user interactions typically involve at least three levels: click/page view, adding to favorites or cart, and purchase.
Similarly, in short-video advertising, user behaviors consist of play duration (e.g., pxs), clicks, and conversions.

A user may interact with the same item at different levels within a short period (e.g., clicking, then favoriting, and finally purchasing).
Suppose we construct interaction sequences based on each user behavior (i.e., next item prediction, where the goal is to predict the next item $v_{n+1}$ given interaction history $[(b_1, v_1), \cdots, (b_n, v_n)]$, and fixed the behavior $b_{n + 1}$).
In that case, models predicting high-level behaviors (such as purchase or conversion) can become overly influenced by recently interacted items (e.g., $v_n, v_{n - 1}$).
However, in a real-world recommender system, user interactions within a short time span often originate from the same set of exposed items.
This implies that, when predicting which items from the current exposure set may lead to high-level behaviors, the model does not actually have access to interactions from that same exposure batch during inference.
As a result, a discrepancy arises between the training input and the inference input.

\begin{figure*}[htbp]
    \centering
    \begin{subfigure}{0.23\textwidth}
        \centering
        \includegraphics[width=\linewidth]{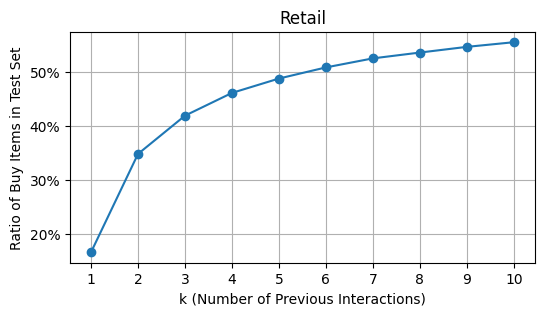}
        \caption{Ratail}
        \label{subfig:a}
    \end{subfigure}
    \hfill
    \begin{subfigure}{0.23\textwidth}
        \centering
        \includegraphics[width=\linewidth]{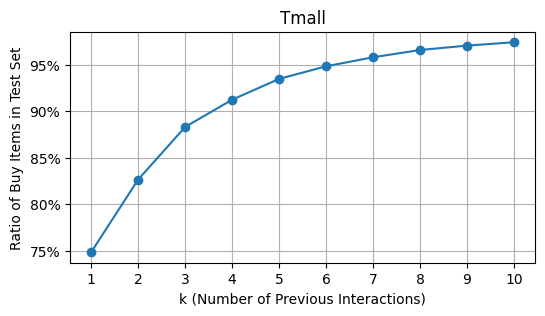}
        \caption{Tmall}
        \label{subfig:b}
    \end{subfigure}
    \hfill
    \begin{subfigure}{0.23\textwidth}
        \centering
        \includegraphics[width=\linewidth]{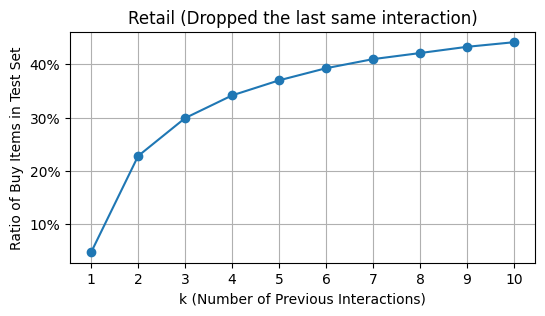}
        \caption{Retail (processed)}
        \label{subfig:c}
    \end{subfigure}
    \hfill
    \begin{subfigure}{0.23\textwidth}
        \centering
        \includegraphics[width=\linewidth]{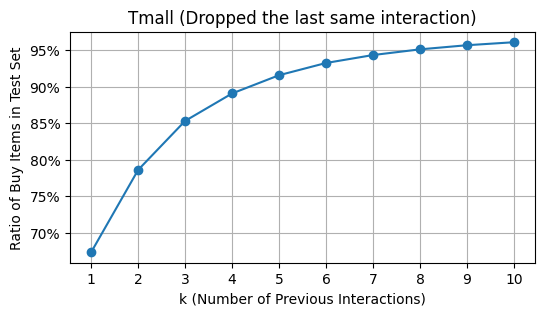}
        \caption{Tmall (processed)}
        \label{subfig:d}
    \end{subfigure}

    \caption{The ratio of items with buy behavior in the test set under the leave-one-out setting for Retail and Tmall. Figure~\ref{subfig:a} and Figure~\ref{subfig:b} show the original data. Figure~\ref{subfig:c} and Figure~\ref{subfig:d} filter the most recent interaction in the user history if it is a low-level behavior for the same target item. It is worth noting that the ratios of $k=5$ and $k=10$ in Figure~\ref{subfig:a} are consistent with the HR@5 and HR@10 reported in MBGen~\cite{liu2024multi}.}
    \Description{}
    \label{fig:duplication}
\end{figure*}

Moreover, in certain multi-behavior scenarios such as e-commerce, a significant proportion of high-level behaviors are preceded by lower-level interactions on the same item within a short interval.
This allows models under the next-item prediction setting to achieve excellent test performance merely by memorizing recent interaction history.
For example, as shown in Figure~\ref{fig:duplication} for the Retail\footnote{https://github.com/anananan116/MBGen/tree/main/data} and Tmall\footnote{https://tianchi.aliyun.com/dataset/140281} datasets under a leave-one-out evaluation, a large portion of purchased items in the test set appear among the most recent $k$ interacted items.
On the Tmall dataset, a model only needs to memorize recent interactions to achieve an over $90\%$ top-5 hit ratio, which is a result clearly inconsistent with practical scenarios.

Therefore, in this paper, we adopt a \textbf{session-wise next item prediction} approach to address the aforementioned inconsistencies.
Specifically, we first partition the interaction sequence of user $u$, denoted as $\mathcal{S}_u = [(b_1, v_1), \cdots, (b_n, v_n)]$, into multiple sessions $\mathcal{S}^\prime_u = [\pmb{S}_1, \cdots, \pmb{S}_m]$ based on temporal criteria (e.g., separate sessions after one day or a fixed period of user inactivity).
$m$ represents the total number of sessions for user $u$.
For each interaction $(b_i, v_i)$ within a session $\pmb{S}_s$, the available historical context is restricted to sessions $\pmb{S}_1$ through $\pmb{S}_{s-1}$, excluding any interactions in $\pmb{S}_s$ that occurred earlier than the current one.

For evaluation, we adopt a strategy similar to leave-one-out for sequential recommendation.
For each user, we hold out the last session as the test set, the second-to-last session as the validation set, and use the remaining interactions for training.
The process can be formulated as, $\mathcal{D}^{\rm{test}} = \{S_{m_u}\}_{u \in \mathcal{U}}, \mathcal{D}^{\rm{val}} = \{S_{m_u - 1}\}_{u \in \mathcal{U}}$.
During the testing phase, given a user’s historical interactions $[\pmb{S}_1, \cdots, \pmb{S}_{m-1}]$ and a fixed target behavior $b$, all items in session $\pmb{S}_m$ with which the user interacted via behavior $b$ are considered positive instances, denoted as $\mathcal{T}_u$.
In addition, we discuss the unification of the training task and testing setup, as detailed in Appendix~\ref{sec:session-wise-training}.


\subsection{Multi-Behavior Sequential Augmentation}
\label{sec:augmentation}

In multi-behavior recommendation scenarios, an inverse relationship often exists between the density of user interactions and the level of behavior.
Higher-level behaviors (e.g., purchase, conversion) typically occur much less frequently than lower-level ones.
This sparsity poses significant challenges for accurately modeling and predicting high-level behaviors.
Although multi-behavior sequential recommendation models often leverage mixed-behavior sequences, attempting to use low-level behaviors to enhance user representation learning, the extreme imbalance in frequency (often by orders of magnitude) remains an issue.
Under constraints on sequence length (e.g., 50 or 100 most recent interactions), the visible history may contain only very few or even zero high-level behaviors, which considerably weakens the model's ability to capture patterns related to these behaviors.
As a result, the model may become overly reliant on low-level interactions, impairing both prediction performance and generalization ability, since the absence of certain low-level signals could disproportionately affect predictions.

To enhance the robustness and generalizability of our sequence-to-sequence model, we introduce a data augmentation strategy specifically designed for multi-behavior user interaction sequences.
For each user $u$ in the training set $\mathcal{D}^{\rm{train}} = \{\mathcal{S}_u\}_{u \in \mathcal{U}}$ with interaction sequence $\mathcal{S}_u$, we apply the following: given a sampling ratio $r$, we randomly discard a proportion $r / L_b$ of interactions for every behavior type $b$ except the highest-level behavior.
$L_b$ denotes the hierarchy level of behavior $b$ (with the lowest level being 1, increasing with behavioral depth).
This results in an augmented sequence $\mathcal{S}_{u, r}$.

For each original user sequence, we generate $x$ augmented versions using ratios $r_i^x = \frac{i}{x + 1}$ for $i = 1, 2, \dots, x$, yielding augmented sequences $\left[\mathcal{S}_{u, r_1^x}, \cdots, \mathcal{S}_{u, r_x^x}\right]$.
These are included in an augmented dataset, $\mathcal{D}^{\rm{aug}}_{x} = \{\mathcal{S}_{u, r_1^x}, \cdots, \mathcal{S}_{u, r_x^x}\}_{u \in \mathcal{U}}$.
The overall enhanced training set becomes $\mathcal{D}_{\rm{aug}_x}^{\rm{train}} = \mathcal{D}^{\rm{train}} \cup \mathcal{D}^{\rm{aug}}_x$.
Notably, the validation set $ \mathcal{D}^{\rm{val}} $ and test set $ \mathcal{D}^{\rm{test}} $ remain unmodified to ensure fair evaluation.

\subsection{Qwen3 MoE Block with Cross-level Behavior Interaction}
\label{sec:cross}

Our proposed architecture comprises three modules in each transformer block: Causal Self-Attention Layer, Cross-level Behavior Interaction Layer, and Position-and-Behavior Aware MoE~\cite{liu2024multi}. 

In the Causal Self-Attention Layer, each item attends only to its preceding items when computing attention weights through a causal mask $M \in \mathbb{R}^{L \times L}$. Assume that the input Query, Key, and Value are $Q_{\text{CA}}, K_{\text{CA}}, V_{\text{CA}} \in \mathbb{R}^{L \times D}$ respectively, where $D$ is the dimension of the feature.
\begin{equation}
    \begin{aligned}
        & \text{Causal-Attention}(Q_{\text{CA}},K_{\text{CA}},V_{\text{CA}}) \\
        = & \text{softmax}\left(\frac{Q_{\text{CA}} \cdot K_{\text{CA}}^T}{\sqrt{D}} \otimes M\right) \cdot V_{\text{CA}}.
    \end{aligned}
\end{equation}
The final attention result passes through the output layer $W_{O_{\text{CA}}} \in \mathbb{R}^{D \times D}$ and serves as the input hidden states of the next module.

Our proposed cross-level behavior interaction layer explicitly models the dependencies between different types of behavior.
To highlight distinctions among behaviors at different levels, we design behavior embedding tables $E_{\mathcal{B}, Q}, E_{\mathcal{B}, K}, E_{\mathcal{B}, V} \in \mathbb{R}^{|\mathcal{B}| \times D}$.
Assume that the behavior of items is $\pmb{B} \in \mathbb{R}^{L}$.
The behavior embeddings are added to $Q_{\text{BA}}$, $K_{\text{BA}}$, and $V_{\text{BA}}$ respectively, according to the behavior sequence $\pmb{B}$.
\begin{equation}
    \begin{aligned}
        Q_{\mathcal{B}} = Q_{\text{BA}} + E_{\mathcal{B}, Q}(\pmb{B}),K_{\mathcal{B}} = K_{\text{BA}} + E_{\mathcal{B}, K}(\pmb{B}),V_{\mathcal{B}} = V_{\text{BA}} + E_{\mathcal{B}, V}(\pmb{B}).
    \end{aligned}
\end{equation}

We then design a behavior-wise mask $M_\mathcal{B} \in \mathbb{R}^{L \times L}$ that restricts each item to interact only with preceding items of lower-level behaviors, explicitly enabling cross-level interactions.
Specifically, for any $i < j$, if the behavior of the $i$-th item is lower than that of the $j$-th item, then $M_\mathcal{B}(i,j) = 1$; otherwise, $M_\mathcal{B}(i,j) = 0$.

\begin{equation}
    \begin{aligned}
        & \text{Behavior-Attention}(Q_{\mathcal{B}},K_{\mathcal{B}},V_{\mathcal{B}}) \\= & \text{softmax}\left(\frac{Q_\mathcal{B} \cdot K^T_\mathcal{B}}{\sqrt{D}} \otimes M_\mathcal{B}\right) \cdot V_\mathcal{B}.
    \end{aligned}
\end{equation}

Finally, we introduce a gating mechanism that learns the importance of cross-level behavior interaction features by generating a weight matrix $G \in \mathbb{R}^{L \times D}$ from the input hidden states $H \in \mathbb{R}^{L \times D}$ through an MLP with SiLU activation function.
\begin{equation}
    \begin{aligned}
        G &= \text{SiLU}(H \cdot W_G), \\
        O &= (\text{Behavior-Attention}(Q_{\mathcal{B}},K_{\mathcal{B}},V_{\mathcal{B}}) \cdot W_{O_{\text{BA}}}) \otimes G,
    \end{aligned}
\end{equation}
where $W_{O_{\text{BA}}} \in \mathbb{R}^{D \times D}$ is the output layer.

The Position-and-Behavior-Aware MoE routes tokens at different positions through fixed expert networks to model the heterogeneity among tokens in the tokenized representation of items, while injecting behavior embedding into the SID tokens before the feed-forward layer to facilitate the fusion of item and behavior information. 
Assume that the input sequence is $S = \left\{ (b^0_i,v^1_i, \cdots, v^l_i) \right\}_{i=1}^{L}$ ($l$ is the number of tokens represented by the semantic IDs of an item), the behavior embedding table is $E_{\mathcal{B}} \in \mathbb{R}^{|\mathcal{B}| \times D}$, and the expert networks are $\phi_j|_{j=0}^{l}$, where $b^0_i \in \mathbb{R}^{D}$ is the behavior token state, $v^j_i \in \mathbb{R}^{D}$ are the SID token states and $o_i^j \in \mathbb{R}^{D}$ is the output state of $i$-th item's $j$-th token.
\begin{equation}
    \begin{aligned}
        o^0_i &= \phi_0(b^0_i), o_i^j = \phi_j\left(\hat v^j_i\right),\\
        \hat v^j_i &= \text{concat}\left(v^j_i,E_{\mathcal{B}}(b_i)\right).
    \end{aligned}
\end{equation}

Finally, the architecture of GAMER is constructed by stacking multiple aforementioned blocks.
It is trained on the sequence $\mathcal{S} = \{x_{1}, x_{2}, \ldots, x_{|\mathcal{S}|}\}$ through a next-token prediction objective:
\begin{align}
    \mathcal{L} = \sum_{t=1}^{T} \log P(x_{t} \mid x_{<t}; \theta),
\end{align}
where $\theta$ denotes the parameters of GAMER.
Note that our decoder-only framework built upon Qwen3 shifts the training granularity from the interaction-level to the \textbf{user-level} by regarding each user as a training sample, thereby improving training efficiency from previous methods with encoder-decoder architectures.

\section{Experiments}

\subsection{Experimental Settings}

\begin{table}[ht]
  \centering
  \caption{Statistics of datasets.}
    \begin{tabular}{c|cccc}
      \toprule
      Dataset & \#User & \#Item & \#Session & \#Inter \\
      \midrule
      ShortVideoAD & 48, 779 & 168, 530 & 1, 874, 719 & 7, 877, 083 \\
      Tmall & 217, 374 & 379, 450 & 856, 756 & 3, 818, 122 \\
      JData & 10, 010 & 17, 100 & 195, 860 & 1, 643, 212 \\
      \bottomrule
    \end{tabular}
  \label{tab:dataset_basic}
\end{table}

\subsubsection{\textbf{Datasets}}
We evaluated GAMER with the following dataset.
Table~\ref{tab:dataset_basic} presents the basic statistical information of datasets.
For all three datasets, users with fewer than three sessions were excluded.
\paragraph{\textbf{ShortVideoAD}}
We utilize a user advertisement interaction dataset, ShortVideoAD, collected from leading short-video platforms.
The dataset encompasses three types of user behaviors: play-throughs exceeding 3 seconds (p3s), clicks, and conversions.
All data were gathered from real user interactions within one week.
User sessions were segmented based on a 15-minute inactivity threshold.
\paragraph{\textbf{Tmall}} Tmall\footnote{https://tianchi.aliyun.com/dataset/140281} is a general e-commerce dataset from Alibaba, containing four types of user behaviors: click, collect, cart, and Alipay.
We selected one week's data and randomly sampled 25\% of the users for experiments.
User sessions were segmented by day.

\paragraph{\textbf{JData}} JData is a general e-commerce dataset from JD\footnote{https://global.jd.com/}, containing five types of user behaviors: page view, click, collect, cart, and purchase.
We randomly sampled 10\% of the users for experiments.
User sessions were segmented based on a 30-minute inactivity threshold.

\subsubsection{\textbf{Compared Methods}}
The selected baseline models are grouped into four categories by their primary technical focus: sequential recommendation, multi-behavior recommendation, multi-behavior sequential recommendation, and multi-behavior generative recommendation.

\paragraph{\textbf{Sequential Recommendation Models}} Follow the settings in \citet{yuan2022multi} and \citet{liu2024multi}, we treat the multi-behavior user interaction sequence as a regular single-behavior user interaction sequence for all sequential recommendation models.
\begin{itemize}
    \item \textbf{Rule-Based} recommends the $k$ most recently interacted unique items from the user's history, ranked in reverse chronological order.
    \item \textbf{GRU4Rec}~\cite{hidasi2015session} is a pioneering model that introduces Gated Recurrent Units (GRUs) to model user interaction sequences for session-based recommendation.
    \item \textbf{SASRec}~\cite{kang2018self} is a canonical model that leverages self-attention mechanisms to capture long-range dependencies in user interaction history.
    \item \textbf{BERT4Rec}~\cite{sun2019bert4rec} adopts a bidirectional Transformer architecture (akin to BERT) to model user behavior sequences.
    It overcomes the limitations of unidirectional models by employing a Cloze (masked language model) training objective, where randomly masked items in the sequence are predicted based on both left and right context.
    \item \textbf{TIGER}~\cite{rajput2023recommender} creates semantic IDs by RQ-VAE for each item and models the sequential recommendation task into a sequential-to-sequential task by predicting the semantic IDs of the next item.
\end{itemize}

\paragraph{\textbf{Multi-Behavior Recommendation Models}}
\begin{itemize}
    \item \textbf{MB-GMN}~\cite{xia2021graph} is a multi-behavior recommendation framework that incorporates a graph meta network, enabling the modeling of multi-behavior patterns within a meta-learning paradigm.
    \item \textbf{S-MBRec}~\cite{gu2022self} is a self-supervised graph model that captures both the differences and commonalities between target and auxiliary behaviors via a contrastive learning task.
\end{itemize}

\paragraph{\textbf{Multi-Behavior Sequential Recommendation Models}}
\begin{itemize}
    \item 
    \textbf{SASRec\textsubscript{B}}~\cite{kang2018self}, and \textbf{BERT4Rec\textsubscript{B}}~\cite{sun2019bert4rec} follow the setting in MBGen~\cite{liu2024multi}, where each distinct type of behavior with an item is treated as a unique token. By remapping the multi-behavior sequence into a unified sequence of these tokens, it transforms the problem into a standard sequential recommendation task for SASRec.
    \item \textbf{PBAT}~\cite{su2023personalized} proposes a Personalized Behavior-Aware Transformer for multi-behavior sequential recommendation that models users' diverse multi-behavior patterns and captures time-evolving behavioral dependencies through a novel fused behavior-aware attention mechanism.
    \item \textbf{MBHT}~\cite{yang2022multi} utilizes a multi-scale transformer with low-rank self-attention and incorporates the global multi-behavior dependency into the hypergraph neural architecture to capture the hierarchical long-range item correlations.
    \item \textbf{MB-STR}~\cite{yuan2022multi} designs a sparse MoE architecture to better leverage multi-behavior supervision and a multi-behavior sequential pattern generator to encode the diverse sequential patterns among multiple behaviors.
\end{itemize}

\paragraph{\textbf{Multi-Behavior Generative Recommendation Models}}
\begin{itemize}
    \item \textbf{TIGER\textsubscript{MB}}~\cite{rajput2023recommender} introduces additional behavior tokens based on the original TIGER vocabulary. In this way, we enable TIGER\textsubscript{MB} to support multiple behavior recommendation tasks.
    \item \textbf{MBGen}~\cite{liu2024multi} tokenizes both the behaviors and items into tokens and constructs one single token sequence. Additionally, it proposes a position-and-behavior-aware transformer with a sparse MoE architecture to better model the heterogeneous nature of token sequences in the generative recommendation. We tested it using chunked IDs (CID) and semantic IDs (SID), respectively.
\end{itemize}

\subsubsection{\textbf{Evaluation Protocols}}
We employ three widely adopted metrics for evaluation.
\begin{itemize}
\item \textbf{Hit Ratio (HR)}: HR@\textit{K} measures whether any target item appears in the Top-\textit{k} generated recommendation list.
\item \textbf{Recall (R)}: R@\textit{K} can be calculated with, $\rm{R}@\textit{K} = \frac{|\rm{TOP}_{\le \textit{K}} \cap \mathcal{T}|}{|\mathcal{T}|}$, where $\rm{TOP}_{\le \textit{K}}$ is the Top-\textit{K} generated recommendation list, and $\mathcal{T}$ is the target item set in the evaluation session.
\item \textbf{Normalized Discounted Cumulative Gain (NDCG)}: The calculation of NDCG@\textit{K} (N@\textit{K}) can be formulated as $\rm{NDCG}@\textit{K} = \dfrac{\rm{DCG}@\textit{K}}{\rm{IDCG}@\textit{K}}$.
DCG@\textit{K} and IDCG@\textit{K} can be calculated as follows,
\begin{equation}
    \begin{aligned}
        \rm{DCG}@\textit{K} & = \sum_{i = 1}^{\textit{K}} \frac{\rm{rel}_i}{\log_2(1 + i)}, \\
        \rm{IDCG}@\textit{K} & = \sum_{i = 1}^{\min(\textit{K}, |\rm{T}|)} \frac{1}{\log_2 (1 + i)},
    \end{aligned}
\end{equation}
where $\rm{rel}_i$ is relevance score of the $i$-th item in the generated recommendation list (i.e., $\rm{rel}_i = 1$ only if $i$-th item is in $\mathcal{T}$).
\end{itemize}

For dataset partitioning, we adopt the session-wise leave-one-out strategy, as described in Section~\ref{sec:session-wise}.
Specifically, for each user, the interactions from the last session are used as the test set, while those from the penultimate session are used as the validation set.
For evaluation within a single session, we fix the user behavior $b$ and treat all items associated with $b$ in that session as positive instances, i.e., the target item set $\mathcal{T}$.
To generate ranking candidates, we apply constrained beam search to all generative recommenders, using a beam size of 20.
The top 10 items retrieved from the full item space are then used for metric computation.

\subsubsection{\textbf{Implementation Details}}
For a fair comparison, all the generative recommenders shared the same vocabulary for item tokenization.
\paragraph{\textbf{Item Tokenizer}}
As for semantic IDs (SIDs), we use a vocabulary size of $(8192 \times 4)$ for all generative recommendation methods.
All the SIDs are obtained by training RQ-VAE based on the modality and interaction features of all advertising videos.
We release the generated SIDs within the ShortVideoAD dataset.
We choose $k = 64$ for generating balanced chunked IDs (CIDs).

\paragraph{\textbf{Sequence-to-Sequence Model}}
We implement our sequence-to-sequence prediction model based on the Qwen3 architecture.
We use a model dimension of 256, an inner dimension of 512 with the SiLU activation function, and 6 heads of dimension 64 in the causal self-attention layer.
The model has 8 decoder layers.
We set the batch size to 4096, the initial learning rate to 0.0005, and trained all the sequence-to-sequence models with the AdamW~\cite{loshchilov2017decoupled} optimizer for 200 epochs.
The learning rate was first linearly warmed up for the first 4\% of the total training steps, followed by a cosine decay schedule to a minimum value of $10^{-6}$.
The models with the lowest validation loss are selected to be loaded at the end of training.

We provide the implementation details of the discriminative baseline in Appendix~\ref{sec:app_impl}.

\begin{table*}
    \centering
    \caption{Performance comparison of different models on the target behavior item prediction task of ShortVideoAD dataset. The best performance is denoted in bold font. We use the \underline{underlined} font to denote the best performance in baseline models.}
    \begin{threeparttable}[c]
    \setlength{\tabcolsep}{3mm}{
        \resizebox{\linewidth}{!}{
        \begin{tabular}{cccccccccc}
            \toprule
            Model Type &Model &HR@1 &HR@5 &HR@10 &R@1 &R@5 &R@10 &N@5 &N@10 \\
            \midrule
            \multirow{5}{*}{\shortstack{Sequential \\ Recommendation}}
            &Rule-Based & 0.0158 & 0.0564 & 0.0841 & 0.0123 & 0.0426 & 0.0626 & 0.0298 & 0.0364 \\
            &GRU4Rec & 0.0256 & 0.0847 & 0.1349 & 0.0175 & 0.0613 & 0.0993 & 0.0438 & 0.0561 \\
            &SASRec & 0.0245 & 0.0894 & 0.1455 & 0.0162 & 0.0622 & 0.1051 & 0.0438 & 0.0577 \\
            &BERT4Rec & 0.0204 & 0.0747 & 0.1212 & 0.0138 & 0.0511 & 0.0849 & 0.0362 & 0.0474 \\
            &TIGER & 0.0249 & 0.0873 & 0.1342 & 0.0176 & 0.0630 & 0.0970 & 0.0447 & 0.0559 \\
            \midrule
            \multirow{2}{*}{\shortstack{Multi-Behavior \\ Recommendation}}
            &MB-GMN & 0.0046 & 0.0108 & 0.0171 & 0.0037 & 0.0093 & 0.0144 & 0.0066 & 0.0083 \\
            &S-MBRec & 0.0013 & 0.0108 & 0.0168 & 0.0013 & 0.0094 & 0.0138 & 0.0051 & 0.0067 \\
            \midrule
            \multirow{5}{*}{\shortstack{Multi-Behavior \\ Sequential \\ Recommendation}}
            &SASRec\textsubscript{B} & 0.0237 & 0.0891 & 0.1471 & 0.0183 & 0.0632 & 0.1059 & 0.0449 & 0.0592 \\
            &BERT4Rec\textsubscript{B} & 0.0017 & 0.0090 & 0.0176 & 0.0007 & 0.0032 & 0.0069 & 0.0028 & 0.0040 \\
            &PBAT & 0.0157 & 0.0683 & 0.1137 & 0.0115 & 0.0494 & 0.0816 & 0.0328 & 0.0436 \\
            &MBHT & \underline{0.0290} & 0.0951 & 0.1526 & 0.0202 & 0.0650 & 0.1073 & 0.0480 & 0.0619 \\
            &MB-STR & \underline{0.0290} & 0.0940 & 0.1515 & \underline{0.0212} & 0.0667 & 0.1104 & 0.0488 & 0.0631 \\
            \midrule
            \multirow{6}{*}{\shortstack{Multi-Behavior \\ Generative \\ Recommendation}}
            & TIGER\textsubscript{MB} (SID) & 0.0283 & 0.0967 & 0.1505 & 0.0207 & 0.0697 & 0.1094 & 0.0497 & 0.0627 \\
            &MBGen (CID) & 0.0288 & 0.1009 & 0.1531 & 0.0198 & 0.0709 & 0.1090 & 0.0510 & 0.0635 \\
            &MBGen (SID) & 0.0276 & \underline{0.1012} & \underline{0.1622} & 0.0202 & \underline{0.0736} & \underline{0.1205} & \underline{0.0518} & \underline{0.0673} \\
            &\textbf{GAMER (CID)} & 0.0381 & 0.1158 & 0.1759 & 0.0272 & 0.0850 & 0.1292 & 0.0620 & 0.0766 \\
            &\textbf{GAMER (SID)} & \textbf{0.0394} & \textbf{0.1280} & \textbf{0.1944} & \textbf{0.0292} & \textbf{0.0966} & \textbf{0.1478} & \textbf{0.0687} & \textbf{0.0856} \\
            &\# Improve & \textbf{+35.86\%} & \textbf{+26.48\%} & \textbf{+19.85\%} & \textbf{+37.74\%} & \textbf{+31.25\%} & \textbf{+22.66\%} & \textbf{+32.63\%} & \textbf{+27.19\%} \\
            \bottomrule
        \end{tabular}
        }
    }
    \end{threeparttable}
    \label{tab:main_exp}
\end{table*}

\begin{table*}
    \centering
    \caption{Performance comparison of different models on the target behavior item prediction task of Tmall and JData dataset.}
    \begin{threeparttable}[c]
    \setlength{\tabcolsep}{3mm}{
        \resizebox{\linewidth}{!}{
        \begin{tabular}{cccccccccc}
            \toprule
            \multirow{2}{*}{Model Type} &\multirow{2}{*}{Model} & \multicolumn{4}{c}{Tmall} & \multicolumn{4}{c}{JData} \\
            \cmidrule(lr){3-6} \cmidrule(lr){7-10}
            & &HR@5 &HR@10 &N@5 &N@10 &HR@5 &HR@10 &N@5 &N@10 \\
            \midrule
            \multirow{2}{*}{\shortstack{Sequential \\ Recommendation}}
            &SASRec & 0.0659 & 0.0768 & 0.0496 & 0.0533 & 0.4609 & 0.5217 & 0.3665 & 0.3860 \\
            &TIGER & \underline{0.5541} & \underline{0.5810} & \underline{0.4584} & \underline{0.4687} & \underline{0.5948} & 0.6509 & \underline{0.4636} & \underline{0.4824} \\
            \midrule
            \multirow{3}{*}{\shortstack{Multi-Behavior \\ Sequential \\ Recommendation}}
            &SASRec\textsubscript{B} & 0.0054 & 0.0065 & 0.0046 & 0.0049 & 0.3696 & 0.4348 & 0.2604 & 0.2818 \\
            &PBAT & 0.2648 & 0.2827 & 0.2121 & 0.2189 & 0.4870 & 0.5348 & 0.3446 & 0.3601 \\
            &MB-STR & 0.3086 & 0.3282 & 0.2561 & 0.2632 & 0.5696 & \underline{0.6522} & 0.4363 & 0.4646 \\
            \midrule
            \multirow{3}{*}{\shortstack{Multi-Behavior \\ Generative \\ Recommendation}}
            &MBGen (CID) & 0.5503 & 0.5721 & 0.4582 & 0.4672 & 0.5647 & 0.6164 & 0.4374 & 0.4520 \\
            &\textbf{GAMER (CID)} & \textbf{0.5628} & \textbf{0.5841} & \textbf{0.4711} & \textbf{0.4793} & \textbf{0.6638} & \textbf{0.7457} & \textbf{0.5302} & \textbf{0.5571} \\
            &\# Improve & \textbf{+1.57\%} & \textbf{+0.53\%} & \textbf{+2.77\%} & \textbf{+2.26\%} & \textbf{+11.60\%} & \textbf{+14.34\%} & \textbf{+14.37\%} & \textbf{+15.49\%} \\
            \bottomrule
        \end{tabular}
        }
    }
    \end{threeparttable}
    \label{tab:main_exp_others}
\end{table*}

\subsection{Results}
\label{sec:results}

\subsubsection{\textbf{Target Behavior Item Prediction}}

For the target behavior item prediction task, we only use user interaction sequences with the last session that contain at least a target behavior (e.g., conversion, purchase) for evaluation.
We present the results on ShortVideoAD in Table~\ref{tab:main_exp}.

Models incorporating user interaction sequences consistently outperform graph-based multi-behavior models, underscoring the importance of sequential information in multi-behavior recommendation scenarios.
Multi-behavior variants of sequential models (e.g., SASRec\textsubscript{B}), which simply treat different behaviors on the same item as distinct items, generally underperform their original versions, indicating that such simplification fails to leverage the hierarchical behavioral characteristics of items and significantly impairs model effectiveness.
The strong performance of MBHT and MB-STR highlights the necessity of explicit behavioral feature modeling in multi-behavior sequential recommendation.
Among generative recommendation models, both the TIGER variant with simple behavior token incorporation and MBGen demonstrate substantial improvements over all discriminative models, revealing the potential of the generative recommendation paradigm.
Our model achieves performance gains exceeding 20\% on most metrics compared to all baselines.
Furthermore, both GAMER and MBGen perform better with Semantic IDs (SID) than with Chunked IDs (CID), validating the effectiveness of incorporating item semantics.
We also observe that the performance ranking among different multi-behavior generative models remains largely consistent under both CID and SID tokenization.
Given that CID-based models offer higher training and inference efficiency due to their smaller vocabulary, using CID for rapid model iteration is highly practical.
Finally, the significant performance gap between all generative models and rule-based methods, which rely solely on user history, indicates that under our current task settings, the models do not merely memorize user history but also generalize well to predict users' other potential interests.

As shown in Table~\ref{tab:main_exp_others}, GAMER achieved the best performance on Tmall and JData.
However, due to the particularity of e-commerce scenarios, our improvement over the baseline is relatively small.

\subsubsection{\textbf{Behavior-Specific Item Prediction}}

\begin{table}
    \centering
    \caption{Performance comparison of different models on the behavior-specific item prediction task.}
    \begin{threeparttable}[c]
        \begin{tabular}{c|cccc}
            \toprule
            Model &HR@5 &HR@10 &N@5 &N@10 \\
            \midrule
            Rule-Based & 0.0650 & 0.1010 & 0.0276 & 0.0339 \\
            GRU4Rec\textsubscript{B} & 0.1046 & 0.1592 & 0.0417 & 0.0515 \\
            SASRec\textsubscript{B} & 0.1102 & 0.1695 & 0.0441 & 0.0550 \\
            BERT4Rec\textsubscript{B} & 0.0727 & 0.1138 & 0.0265 & 0.0326 \\
            PBAT & 0.0834 & 0.1334 & 0.0318 & 0.0404 \\
            MB-STR & 0.1017 & 0.1595 & 0.0414 & 0.0518 \\
            \midrule
            TIGER\textsubscript{MB} (SID) & 0.1087 & 0.1627 & 0.0443 & 0.0540 \\
            MBGen (CID) & 0.1108 & 0.1627 & 0.0453 & 0.0542 \\
            MBGen (SID) & \underline{0.1179} & \underline{0.1774} & \underline{0.0493} & \underline{0.0603} \\
            \midrule
            \textbf{GAMER (CID)} & 0.1277 & 0.1851 & 0.0543 & 0.0646 \\
            \textbf{GAMER (SID)} & \textbf{0.1443} & \textbf{0.2129} & \textbf{0.0621} & \textbf{0.0753} \\
            \bottomrule
        \end{tabular}
    \end{threeparttable}
    \label{tab:behavior_specific}
\end{table}

We compare GAMER with various multi-behavior sequential recommendation models under the behavior-specific ttem prediction setting, where the last session of each user sequence is held out for testing.
In this setting, each behavior type within these sessions is evaluated separately, and models are provided with the user's historical sequence along with the specific behavior type to be predicted.

As shown in Table~\ref{tab:behavior_specific}, generative methods maintain their performance advantage over discriminative approaches.
GAMER consistently achieves the best performance among all the metrics.
Notably, models utilizing SID tokenization consistently outperform those using CID on behavior-specific item prediction, further validating the benefit of semantic representations.

\subsection{Ablation Study}

All our ablation experiments were conducted on the ShortVideoAD dataset.

\subsubsection{\textbf{Sequential Augmentation Ablation}}

\begin{figure*}[ht]
    \includegraphics[width=0.85\textwidth]{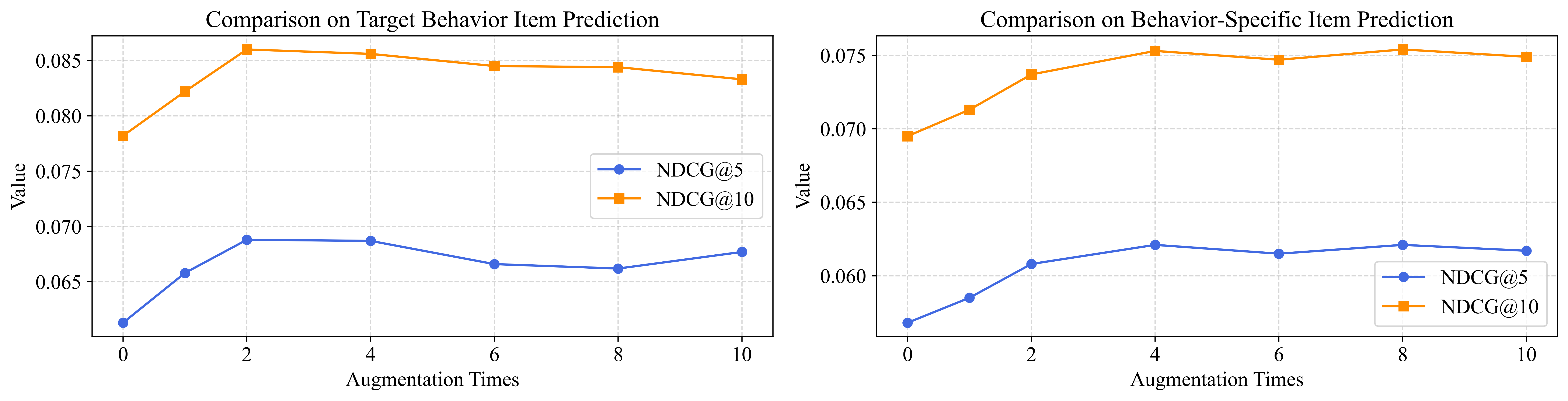}
    \caption{The comparison of different augmentation times on both target behavior item prediction and behavior-specific item prediction tasks. We uniformly use the same SIDs for item tokenization to ensure fairness in comparison.}
    \Description{}
    \label{fig:augmentation}
\end{figure*}

We conduct an ablation study on sequential augmentation times with unified SID tokenization.
Experiments are performed on both target behavior item prediction and behavior-specific item prediction tasks with different augmentation times $x = 0, 1, 2, 4, 6, 8, 10$, where $x = 0$ indicates no augmentation.

As illustrated in Figure~\ref{fig:augmentation}, all models with sequence augmentation demonstrate superior NDCG performance compared to the non-augmented baseline, validating the effectiveness of behavior hierarchy-based sequence augmentation for multi-behavior recommendation.
Overall, performance remains relatively stable across augmentation times ranging from 2 to 10.
Specifically, for target behavior item prediction, the model achieves optimal results with $x = 2$ and $ 4$, while for behavior-specific item prediction, the best performance is observed at $x = 4$ and $ 8$.
Considering these results, we adopt $4\times$ augmentation for all experiments reported in Section~\ref{sec:results} to balance performance across both tasks.

\subsubsection{\textbf{Sequence-to-Sequence Model Architecture}}

\begin{table}
    \centering
    \caption{Performance comparison of different sequence-to-sequence model architectures with our proposed multi-behavior sequential augmentation on the target behavior item prediction task. We uniformly use the same SIDs for item tokenization to ensure fairness in comparison.}
    \begin{threeparttable}[c]
        \begin{tabular}{l|cccc}
            \toprule
            Architecture & HR@5 & HR@10 & N@5 & N@10 \\
            \midrule
            Qwen3 & 0.1008 & 0.1580 & 0.0542 & 0.0689 \\
            \quad + $4\times$ aug. & \textbf{0.1111} & \textbf{0.1672} & \textbf{0.0569} & \textbf{0.0714} \\
            \midrule
            GAMER & 0.1146 & 0.1812 & 0.0613 & 0.0782 \\
            \quad + $4\times$ aug. & \textbf{0.1280} & \textbf{0.1944} & \textbf{0.0687} & \textbf{0.0856} \\
            \midrule
            PBATrans. & 0.1012 & 0.1622 & 0.0518 & 0.0673 \\
            \quad + $1\times$ aug. & \textbf{0.1122} & \textbf{0.1699} & \textbf{0.0579} & \textbf{0.0730} \\
            \bottomrule
        \end{tabular}
    \end{threeparttable}
    \label{tab:ablation_arch}
\end{table}

We further conduct an ablation analysis on different sequence-to-sequence model architectures.
As shown in Table~\ref{tab:ablation_arch}, we evaluate the original Qwen3 architecture, GAMER with multi-level behavior modeling, and the PBATransformer proposed in MBGen~\cite{liu2024multi}.
Each architecture is tested on both the original training set and with $1\times$ or $4\times$ sequence augmentation.
Notably, since PBATransformer's encoder-decoder structure only supports interaction-level training, which differs significantly in training efficiency from the user-level training used in Qwen3-based models (as discussed in Section~\ref{sec:training_cost}), high augmentation multipliers would substantially increase experimental cost.
Thus, we only test it with $1\times$ sequence augmentation.

Comparisons between augmented and non-augmented performances demonstrate that our sequence augmentation method consistently improves performance across all sequence-to-sequence architectures.
Furthermore, the proposed multi-level behavior modeling yields significant performance gains over the original Qwen3 structure, validating its effectiveness in helping the model understand rich hierarchical behavior patterns.

\subsection{Training Cost Analysis}
\label{sec:training_cost}

\begin{table}[ht]
  \centering
  \caption{The training cost for different generative recommendation methods using SIDs for item tokenization. All experiments were conducted using 8 Nvidia A800 GPUs.}
    \begin{tabular}{lc|cc}
      \toprule
      Model & \#Parameters & Training Time & \#Epochs\\
      \midrule
      MBGen & \multirow{2}{*}{26.80M} & $\sim 100 \rm{h}$ & 154 \\
      \quad + $1\times$ aug. & & $\sim 151 \rm{h}$ & 171 \\
      \midrule
      GAMER & \multirow{3}{*}{28.87M} & $\sim 13 \rm{h}$ & 200 \\
      \quad + $4\times$ aug. & & $\sim 12 \rm{h}$ & 94 \\
      \quad + $10\times$ aug. & & $\sim 12 \rm{h}$ & 56 \\
      \bottomrule
    \end{tabular}
  \label{tab:training_cost}
\end{table}

We compare the training efficiency between MBGen~\cite{liu2024multi} (the encoder-decoder architecture) and GAMER (the decoder-only architecture).
As shown in Table~\ref{tab:training_cost}, our approach demonstrates significant training efficiency advantages over MBGen while maintaining comparable parameter counts, primarily due to the inherent compatibility of the decoder-only architecture with user-level training.
Furthermore, although sequence augmentation increases the training set size, it also accelerates model convergence (requiring fewer epochs).
Consequently, $4\sim 10 \times$ sequence augmentation does not substantially increase overall training time, highlighting the additional advantage of our augmentation method.




\section{Conclusion}

In this work, we propose \textbf{GAMER}, a multi-behavior generative recommendation approach incorporating multi-behavior sequential augmentation and hierarchical behavior modeling.
We collect and release \textbf{ShortVideoAD}, a real-world dataset containing three hierarchical user behaviors in short-video advertising with pretrained semantic IDs for generative recommendation.
For this task and dataset, we design a decoder-only architecture.
Extensive experiments demonstrate the effectiveness and robustness of our proposed GAMER.
Building upon this work, promising future directions include exploring diverse data augmentation strategies and advancing multi-behavior interaction modeling within the multi-behavior generative recommendation paradigm.



\bibliographystyle{ACM-Reference-Format}
\balance
\bibliography{main}

\newpage
\appendix

\section{Baseline Implementation Details}
\label{sec:app_impl}

We implement all the multi-behavior graph-based recommendation methods based on SSLRec~\cite{ren2024sslrec}.
For the sequential recommendation models, we primarily refer to the implementation of RecBole~\cite{recbole[1.0]}.
Regarding other multi-behavior sequential recommendation models, we consulted the original code implementations provided in the respective baseline papers.

To facilitate future research, we have open-sourced all baseline implementations for both single-behavior and multi-behavior sequential recommendation.
In our implementation, all sequential recommendation models employ random negative sampling for items in the validation set.
Under the condition that the total number of positive samples and negative samples in each validation session is 1,000, the models perform scoring and ranking.
We use the NDCG@10 results on the validation set of the target behavior item prediction task as the criterion for early stopping (with a patience of 10).
The model checkpoint that achieves the best performance on the validation set is ultimately used for the final evaluation on the test set.

For hyperparameter settings, we uniformly adopted a learning rate of 0.001 with the AdamW optimizer.
The batch size for each model was adjusted dynamically based on GPU memory constraints, with a maximum limit of 4096.
For models trained at the interaction level, we set the maximum historical sequence length to 50, while for those trained at the user level, the maximum length was set to 100.
Due to MBHT 's~\cite{yang2022multi} requirement that the maximum history length relates to its unique multi-scale hyperparameter, we therefore selected a maximum history length of 59 for compatibility.
Additionally, following its original implementation, we only trained on interactions corresponding to the target behavior.
Model-specific hyperparameters are provided in the \verb|config| directory within our code.

\section{Model Robustness Analysis}
\label{sec:robust}

In this section, we further experimentally validate the improvement in model robustness achieved through sequence augmentation.
Following a procedure similar to that described in Section~\ref{sec:augmentation}, we modified the original test setup by randomly removing a proportion $r$ of low-level behaviors from the test sequences.
Under the condition that model inputs were subjected to dropout, we evaluated the performance of TIGER\textsubscript{MB}, MBGen, and our proposed method on the ShortVideoAD dataset for the Target Behavior Item Prediction task.
Additionally, we tested model performance after removing all target items that appeared in the test sessions from the input historical sequences.
As shown in Table~\ref{tab:robustness_analysis}, our method consistently outperforms the baselines under different dropout ratios.
Notably, even when $r=1$, i.e., all the lowest-level interactions were removed, our approach still achieved considerable performance and exhibited more pronounced improvements compared to the two baselines.

\begin{figure}[ht]
    \includegraphics[width=\linewidth]{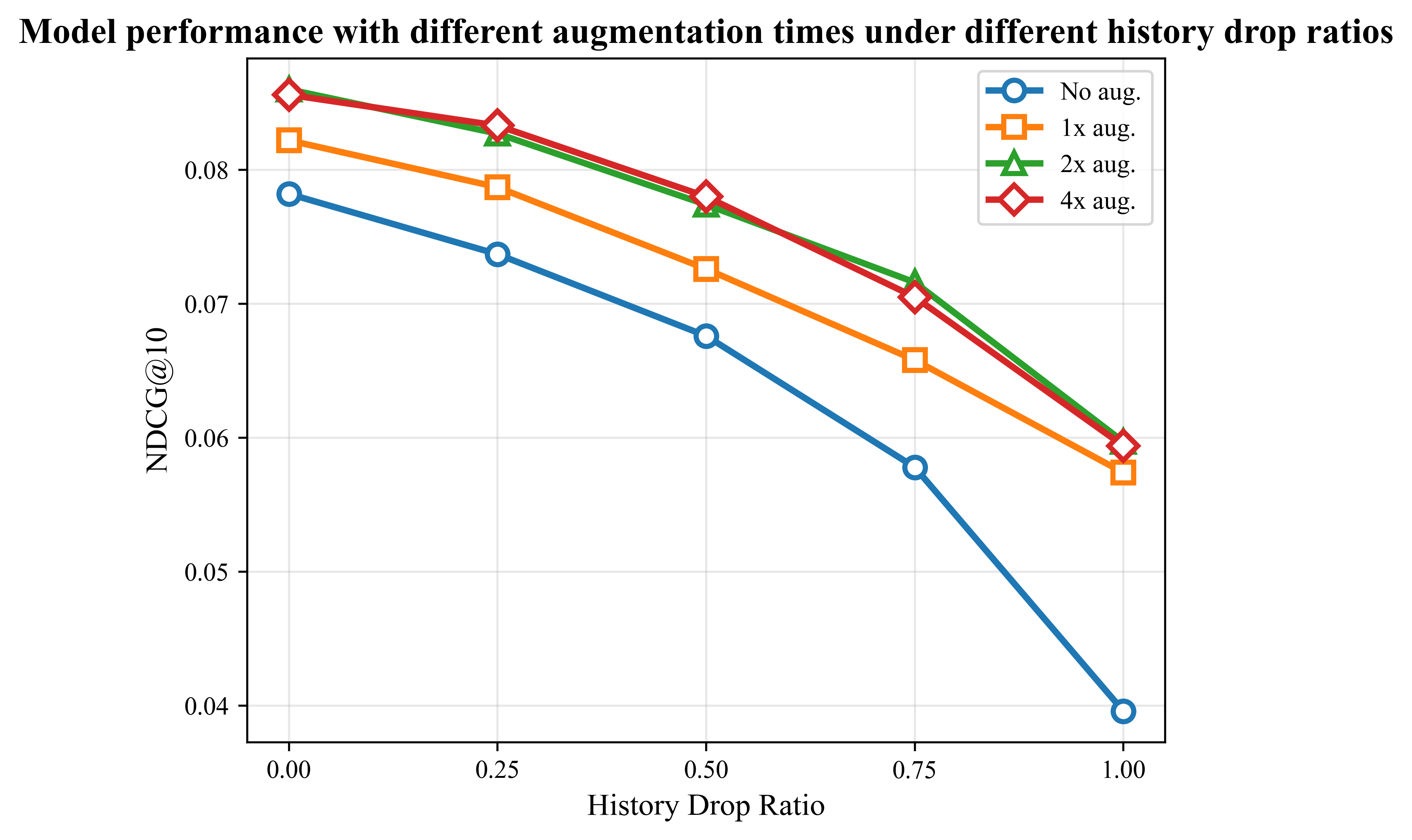}
    \caption{Robustness analysis on sequential augmentation times.}
    \Description{}
    \label{fig:robustness}
\end{figure}

We also conducted experiments on model robustness under different augmentation multipliers.
As illustrated in Figure~\ref{fig:robustness}, models trained with sequence augmentation show a clear performance gap compared to those without augmentation.
Moreover, when the augmentation multiplier $x$ is set to 2 or 4, model performance shows little difference, with both cases yielding improvements over the setting where $x=1$.

\begin{table}[ht]
  \centering
  \caption{Robustness analysis on the target behavior item prediction task.}
  \setlength{\tabcolsep}{1mm}{
    \begin{tabular}{lc|cccc}
      \toprule
       & Model & HR@5 & HR@10 & N@5 & N@10\\
      \midrule
       \multirow{4}{*}{w/o dropout}& TIGER\textsubscript{MB}  &0.0967 &0.1505 &0.0497 &0.0627 \\
       & MBGen  &\underline{0.1012} &\underline{0.1622} &\underline{0.0518} &\underline{0.0673} \\
       & GAMER  &\textbf{0.1280} &\textbf{0.1944} &\textbf{0.0687} &\textbf{0.0856} \\
       &\# Improve & \textbf{+26.48\%} & \textbf{+19.85\%} & \textbf{+32.63\%} & \textbf{+27.19\%}\\
      \midrule
       \multirow{4}{*}{dropout 0.25}& TIGER\textsubscript{MB}  &0.0927 &0.1489 &0.0477 &0.0618 \\
       & MBGen  &\underline{0.0984} &\underline{0.1540} &\underline{0.0508} &\underline{0.0648} \\
       & GAMER  &\textbf{0.1240} &\textbf{0.1930} &\textbf{0.0657} &\textbf{0.0833} \\
       &\# Improve & \textbf{+26.02\%} & \textbf{+25.32\%} & \textbf{+29.33\%} & \textbf{+28.55\%}\\
      \midrule
       \multirow{4}{*}{dropout 0.5} & TIGER\textsubscript{MB}  &0.0911 &0.1439 &0.0449 &0.0582 \\
       & MBGen  &\underline{0.0961} &\underline{0.1515} &\underline{0.0490} &\underline{0.0626} \\
       & GAMER  &\textbf{0.1162} &\textbf{0.1827} &\textbf{0.0611} &\textbf{0.0780} \\
       &\# Improve & \textbf{+20.92\%} & \textbf{+20.59\%} & \textbf{+24.69\%} & \textbf{+24.60\%}\\
      \midrule
       \multirow{4}{*}{dropout 0.75} & TIGER\textsubscript{MB}  &0.0827 &0.1291 &0.0397 &0.0510 \\
       & MBGen  &\underline{0.0890} &\underline{0.1407} &\underline{0.0436} &\underline{0.0561} \\
       & GAMER  &\textbf{0.1100} &\textbf{0.1692} &\textbf{0.0557} &\textbf{0.0705} \\
       &\# Improve & \textbf{+23.60\%} & \textbf{+20.26\%} & \textbf{+27.75\%} & \textbf{+25.67\%}\\
      \midrule
       \multirow{4}{*}{dropout 1.0} & TIGER\textsubscript{MB}  &0.0542 &0.0870 &0.0256 &0.0333 \\
       & MBGen  &\underline{0.0560} &\underline{0.0872} &\underline{0.0272} &\underline{0.0347} \\
       & GAMER  &\textbf{0.0919} &\textbf{0.1464} &\textbf{0.0461} &\textbf{0.0594}\\
       &\# Improve & \textbf{+64.11\%} & \textbf{+67.89\%} & \textbf{+69.49\%} & \textbf{+71.18\%}\\
     \midrule
       \multirow{4}{*}{dropout gt} & TIGER\textsubscript{MB}  &0.0889 &0.1392 &0.0444 &0.0567 \\
       & MBGen  &\underline{0.0911} &\underline{0.1526} &\underline{0.0457} &\underline{0.0614} \\
       & GAMER  &\textbf{0.1113} &\textbf{0.1758} &\textbf{0.0570} &\textbf{0.0733} \\
       &\# Improve & \textbf{+22.17\%} & \textbf{+15.20\%} & \textbf{+24.73\%} & \textbf{+19.38\%}\\
      \bottomrule
    \end{tabular}}
  \label{tab:robustness_analysis}
\end{table}

\section{Session-wise Training}
\label{sec:session-wise-training}

\begin{figure*}[ht]
    \includegraphics[width=\textwidth]{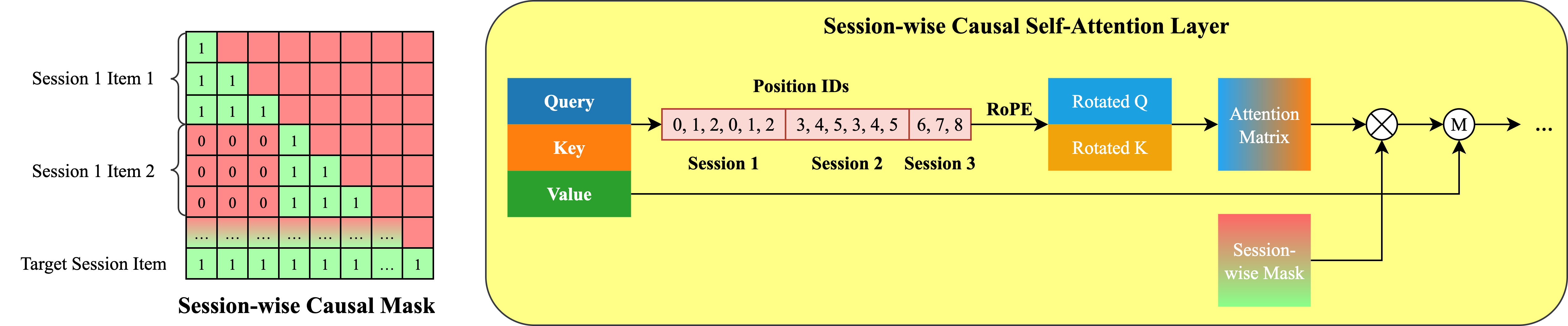}
    \caption{Session-wise Causal Self-Attention Layer.}
    \Description{Session}
    \label{fig:session-wise-attention}
\end{figure*}

\begin{table*}
    \centering
    \caption{Performance comparison of different sequence-to-sequence model architectures with and without session-wise causal self-attention layer design on ShortVideoAD under target behavior item prediction task.}
    \begin{threeparttable}[c]
        \begin{tabular}{c|lcccccccc}
            \toprule
            Augmentation Times & Architecture & HR@1 & HR@5 & HR@10 & R@1 & R@5 & R@10 & N@5 & N@10 \\
            \midrule
            \multirow{4}{*}{No augmentation} & Qwen3 & \textbf{0.0349} & 0.1008 & 0.1580 & \textbf{0.0254} & 0.0724 & \textbf{0.1174} & \textbf{0.0542} & \textbf{0.0689} \\
            & \quad w/ session-wise & 0.0298 & \textbf{0.1034} & \textbf{0.1590} & 0.0208 & \textbf{0.0728} & 0.1150 & 0.0520 & 0.0657 \\
            \cmidrule{2-10}
            & GAMER & \textbf{0.0361} & \textbf{0.1146} & \textbf{0.1812} & \textbf{0.0277} & \textbf{0.0846} & \textbf{0.1360} & \textbf{0.0613} & \textbf{0.0782} \\
            & \quad w/ session-wise & 0.0351 & 0.1112 & 0.1783 & 0.0253 & 0.0805 & 0.1334 & 0.0585 & 0.0759 \\
            \midrule
            \multirow{4}{*}{+ $4\times$ augmentation} & Qwen3 & 0.0318 & 0.1111 & 0.1672 & 0.0239 & 0.0792 & 0.1232 & 0.0569 & 0.0714 \\
            & \quad w/ session-wise & \textbf{0.0344} & \textbf{0.1122} & \textbf{0.1737} & \textbf{0.0243} & \textbf{0.0796} & \textbf{0.1262} & \textbf{0.0580} & \textbf{0.0733} \\
            \cmidrule{2-10}
            & GAMER & \textbf{0.0394} & \textbf{0.1280} & \textbf{0.1944} & \textbf{0.0292} & \textbf{0.0966} & \textbf{0.1478} & \textbf{0.0687} & \textbf{0.0856} \\
            & \quad w/ session-wise & 0.0367 & 0.1237 & 0.1884 & 0.0262 & 0.0904 & 0.1388 & 0.0645 & 0.0804 \\
            \bottomrule
        \end{tabular}
    \end{threeparttable}
    \label{tab:session-wise}
\end{table*}

To unify the training and inference settings, we propose the Session-wise Causal Self-Attention Layer, which is based on the original Causal Self-Attention Layer.

As illustrated in Figure~\ref{fig:session-wise-attention}, we replace the Rotary Position Embedding (RoPE)~\cite{su2024roformer} positional encoding of each token with session-wise position IDs $P_{\mathcal{S}}$, where multiple items within the same session share identical position IDs. 
\begin{equation}
    \begin{aligned}
        Q_{\mathcal{S}} = Q \cdot R_{\Theta}(P_{\mathcal{S}}),K_{\mathcal{S}} = K \cdot R_{\Theta}(P_{\mathcal{S}})
    \end{aligned}
\end{equation}
In addition, to preserve the causal structure at the session level, we design a session-wise causal mask $M_\mathcal{S} \in \mathbb{R}^{L \times L}$, which restricts each token to attend only to items within preceding sessions (excluding the current session).
For any $i < j$, if the session of the $i$-th item is lower than precedes that of the $j$-th item, then $M_\mathcal{S}(i,j) = 1$; otherwise, $M_\mathcal{S}(i,j) = 0$.
\begin{align}
    &\text{Session-Attention}(Q_{\mathcal{S}},K_{\mathcal{S}},V) \\&= \text{softmax}\left(\frac{Q_{\mathcal{S}} \cdot K_{\mathcal{S}}^T}{\sqrt{D}} \otimes M_\mathcal{S}\right) \cdot V
\end{align}

This design ensures consistency between training and inference under the session-wise next-item prediction setting. However, this design also sacrifices the intra-session order information. As shown in Table~\ref{tab:session-wise}, our method exhibits a slight degradation across all evaluation metrics. In future work, we need to design a novel positional encoding that enables each item to be aware of the intra-session order of items within preceding sessions.
\section{GAMER for Ranking Scenarios}

While NTP-based methods excel in dense interaction sequences, they struggle in ad ranking scenarios where target conversion behaviors (e.g., activation, purchase) are extremely sparse. 
To address this, we adapt GAMER to the ranking setting by reframing the task as behavior prediction conditioned on a given item—enabling the model to leverage multi-level user behaviors and effectively estimate sparse conversion probabilities within a generative framework.

%
In the ranking setting, the core objective is to accurately estimate the probability of a specific user behavior given a candidate item.
Inspired by HSTU~\cite{zhai2024actions}, we restructure each user’s session-aware historical interactions—originally introduced in Section~\ref{sec:session-wise}—into the sequence $\mathcal{S}_u = [(v_1, b_1), \cdots, (v_n, b_n)]$, where each item $v_i$ precedes its associated behavior $b_i$. 
Recognizing the substantial semantic disparity between behavior tokens and item semantic tokens, we adopt a vocabulary separation strategy to enhance the precision of behavior prediction.
Specifically, we maintain two independent vocabularies: one for item semantic ID tokens and another for behavior tokens.

The backbone architecture of GAMER remains unchanged. 
However, during prediction, we employ two separate output heads—one dedicated to predicting item semantic IDs and the other to predicting behavior tokens. 
At inference time, the behavior label of each candidate item is replaced with a special \textbf{[MASK]} token, and the model performs behavior prediction conditioned on this masked input.

\subsection{Offline Experimental Results}

\begin{table}[ht]
  \centering
  \caption{Supplementary Dataset Statistics.}
  \scalebox{0.9}{
    \begin{tabular}{c|cccc}
      \toprule
      Dataset & \#User & \#Item & \#Session & \#Inter \\
      \midrule
      ShortVideoAD & 48, 779 & 168, 530 & 1, 874, 719 & 7, 877, 083 \\
      ShortVideoAD\textsubscript{big} & 553, 694 & 1, 275, 632 & 14, 688, 254 & 62, 683, 353 \\
      \bottomrule
    \end{tabular}
    }
  \label{tab:Supplementary Dataset Statistics}
\end{table}

For offline experiments, on the data side, given that advertising scenarios place strong emphasis on user conversion—and that conversion events are extremely sparse—we adopt the same data collection strategy as ShortVideoAD to construct a significantly larger dataset, dubbed ShortVideoAD\textsubscript{big}, which is nearly ten times the size of the original. 
This expansion ensures sufficient behavioral signals for reliable training and evaluation. %
Notably, we retain only exposure and conversion events to better align with real-world advertising conditions.
Statistics of ShortVideoAD\textsubscript{big} are summarized in Table~\ref{tab:Supplementary Dataset Statistics}.

As our baseline, we use a production-grade discriminative model currently deployed on leading short-video platforms. We denote this model as DLRM (Deep Learning Recommendation Model).
This model employs a 4-layer Transformer architecture: two encoder layers process the user’s historical behavior sequence, while two decoder layers handle the candidate item for CTR estimation.

We adopt AUROC as the evaluation metric to assess the model’s ability to discriminate conversion events.
The average results over five runs are reported in Table~\ref{tab:ranking experiments}.

\begin{table}
    \centering
    \caption{Offline experimental results in the ranking scenario.}
        \begin{tabular}{c|c}
            \toprule
            Model & AUROC \\
            \midrule
            DLRM & 0.7975 \\
            GAMER & \textbf{0.8353}\\
            \bottomrule
        \end{tabular}
    \label{tab:ranking experiments}
\end{table}

Experimental results show that GAMER outperforms the state-of-the-art discriminative model currently deployed in production and demonstrates exceptional performance in behavior prediction. 
This is attributed to GAMER’s joint modeling of the \textlangle item, behavior\textrangle\ probability distribution and the Cross-level Behavior Interaction Layer, which effectively enhances prediction accuracy for sparse user behaviors.

\subsection{Online Performance}
We deploy GAMER on a leading short-video platform serving hundreds of millions of daily active users.
To evaluate its online performance, we conduct a one-week A/B test using a 10.0\% traffic slice, comparing GAMER against the current production model. 
The results are reported in Table~\ref{tab:online experiments}.
Our primary evaluation metrics are Conversion Rate (CVR), which reflects the total number of user conversions, and effective Cost Per Mille (eCPM)—a proxy for platform advertising revenue.

\begin{table}
    \centering
    \caption{The relative improvement of our online A/B testing on a short-video advertising scenario.}
        \begin{tabular}{c|c}
            \toprule
            Online Metrics & GAMER \\
            \midrule
             CVR & +2.5\%\\
            eCPM & +1.8\% \\
            \bottomrule
        \end{tabular}
    \label{tab:online experiments}
\end{table}

Experimental results show that GAMER achieves a +2.5\% increase in total conversions—gains and a  +1.8\% improvement in eCPM that are highly significant in the context of activation scenarios.
We regard these findings as strong evidence of GAMER’s practical effectiveness and view generative modeling for ad ranking as a promising direction for future research. 
In particular, we plan to further investigate scaling laws in ranking scenarios and explore the unification of retrieval and ranking within a single generative framework.

\end{document}